\documentstyle[12pt]{article}

\catcode`\@=11
\long\def\@makefntext#1{
\protect\noindent \hbox to 3.2pt {\hskip-.9pt
$^{{\ninerm\@thefnmark}}$\hfil}#1\hfill}		

\def\@makefnmark{\hbox to 0pt{$^{\@thefnmark}$\hss}}  
\def\ps@myheadings{\let\@mkboth\@gobbletwo
\def\@oddhead{\hbox{}
\rightmark\hfil\ninerm\thepage}
\def\@oddfoot{}\def\@evenhead{\ninerm\thepage\hfil
\leftmark\hbox{}}\def\@evenfoot{}
\def\sectionmark##1{}\def\subsectionmark##1{}}

\renewcommand{\thefootnote}{\fnsymbol{footnote}}

\newcounter{sectionc}\newcounter{subsectionc}\newcounter{subsubsectionc}
\renewcommand{\section}[1] {\vspace*{0.6cm}\addtocounter{sectionc}{1}
\setcounter{subsectionc}{0}\setcounter{subsubsectionc}{0}\noindent
	{\normalsize\bf\thesectionc. #1}\par\vspace*{0.4cm}}
\renewcommand{\subsection}[1] {\vspace*{0.6cm}\addtocounter{subsectionc}{1}
	\setcounter{subsubsectionc}{0}\noindent
	{\normalsize\it\thesectionc.\thesubsectionc. #1}\par\vspace*{0.4cm}}
\renewcommand{\subsubsection}[1]
{\vspace*{0.6cm}\addtocounter{subsubsectionc}{1}
	\noindent {\normalsize\rm\thesectionc.\thesubsectionc.\thesubsubsectionc.
	#1}\par\vspace*{0.4cm}}

\newcounter{appendixc}
\newcounter{subappendixc}[appendixc]
\newcounter{subsubappendixc}[subappendixc]

\renewcommand{\appendix}[1] {\vspace*{0.6cm}
        \refstepcounter{appendixc}
        \setcounter{figure}{0}
        \setcounter{table}{0}
        \setcounter{equation}{0}
        \renewcommand{\thefigure}{\Alph{appendixc}.\arabic{figure}}
        \renewcommand{\thetable}{\Alph{appendixc}.\arabic{table}}
        \renewcommand{\theappendixc}{\Alph{appendixc}}
        \renewcommand{\theequation}{\Alph{appendixc}.\arabic{equation}}
        \noindent{\bf Appendix \theappendixc #1}\par\vspace*{0.4cm}}

\newcounter{itemlistc}
\newcounter{romanlistc}
\newcounter{alphlistc}
\newcounter{arabiclistc}

\newcommand{\fcaption}[1]{
        \refstepcounter{figure}
        \setbox\@tempboxa = \hbox{\footnotesize Fig.~\thefigure. #1}
        \ifdim \wd\@tempboxa > 6in
           {\begin{center}
        \parbox{6in}{\footnotesize\baselineskip=12pt Fig.~\thefigure. #1}
            \end{center}}
        \else
             {\begin{center}
             {\footnotesize Fig.~\thefigure. #1}
              \end{center}}
        \fi}

\newcommand{\tcaption}[1]{
        \refstepcounter{table}
        \setbox\@tempboxa = \hbox{\footnotesize Table~\thetable. #1}
        \ifdim \wd\@tempboxa > 6in
           {\begin{center}
        \parbox{6in}{\footnotesize\baselineskip=12pt Table~\thetable. #1}
            \end{center}}
        \else
             {\begin{center}
             {\footnotesize Table~\thetable. #1}
              \end{center}}
        \fi}

\def\@citex[#1]#2{\if@filesw\immediate\write\@auxout
	{\string\citation{#2}}\fi
\def\@citea{}\@cite{\@for\@citeb:=#2\do
	{\@citea\def\@citea{,}\@ifundefined
	{b@\@citeb}{{\bf ?}\@warning
	{Citation `\@citeb' on page \thepage \space undefined}}
	{\csname b@\@citeb\endcsname}}}{#1}}

\newif\if@cghi
\def\cite{\@cghitrue\@ifnextchar [{\@tempswatrue
	\@citex}{\@tempswafalse\@citex[]}}
\def\citelow{\@cghifalse\@ifnextchar [{\@tempswatrue
	\@citex}{\@tempswafalse\@citex[]}}
\def\@cite#1#2{{$\null^{#1}$\if@tempswa\typeout
	{IJCGA warning: optional citation argument
	ignored: `#2'} \fi}}

 1
 1
 1

\font\ninerm=cmr9

\begin{document}

\baselineskip=20pt

\centerline{\normalsize\bf NUCLEAR PHYSICS ---}
\centerline{\normalsize\bf AT THE FRONTIERS OF KNOWLEDGE}

\vspace*{0.5cm}

\baselineskip=13pt

\centerline{\footnotesize HERMAN FESHBACH}
\centerline{\footnotesize\it Center for Theoretical Physics}
\centerline{\footnotesize\it Massachusetts Institute of Technology}
\centerline{\footnotesize\it 77 Massachusetts Avenue}
\centerline{\footnotesize\it Cambridge, MA  02139, USA}
\centerline{\footnotesize E-mail: herman@mitlns.mit.edu}

\vspace*{0.5cm}


\vspace*{0.5cm}

\normalsize\baselineskip=15pt plus 0.5pt minus 0.5pt
\setcounter{footnote}{0}
\renewcommand{\thefootnote}{\alph{footnote}}


Nuclear physics has been and will be a major factor in
science and technology.  It makes unique and important contributions to
medicine, to industry and to other sciences.  Interaction with other
physics has been strong.  Astrophysics and mesoscopic physics are notable
examples.  This is illustrated by Figure~1.  But what I shall talk
about today are labeled {\bf``universals"}.  This refers to results which
transcend the limits of a given subject providing fundamental principles which
inform not only nuclear physics but all of science.  It is the universals
which take nuclear physics research from the parochial albeit fascinating
studies of nuclear reactions and structure to the development and formulation
of concepts of significance for all of the physical sciences.

Fundamentally nuclear
physics research is concerned with the physics of systems consisting of a small
number of interacting particles.  By small number we mean much less than the
$10^{23}$ particles of condensed matter but greater than one.  What have we
discovered about such systems in the last fifty years?  When I look back over
that period, I am struck by the advances that were made, and this in spite of
the fact that certainly at the beginning and for some time longer we only
knew that the nucleon-nucleon force was strong, short range, spin and isospin
dependent.  It was in nuclear physics that the concept of internal symmetry,
e.g. isospin was introduced.  Some of the results which were uncovered
include most importantly the existence of collective motion.  This includes
the existence of a mean field which leads to the shell model, the deformed
nucleus leading to rotational bands and most recently to superdeformed
nuclei (Figure~2),
the doorway states such as the giant electromagnetic resonances, the isobar
analog states, the Gamow-Teller resonances all which seem to be connected with
a symmetry of the nuclear Hamiltonian fission isomers and nuclear molecules
reflecting the properties of highly deformed nuclei.  The whole panorama of
nuclear reactions has been observed, ranging from direct reactions, to
compound nuclear formation and including, in the case of heavy ions, fusion
and fission and deep inelastic scattering.  Theoretically, descriptions of
nuclear structure and the giant resonances have been formulated and
importantly it has been possible to obtain the properties of nuclei starting
from the nucleon-nucleon force.  And a formalism has been developed which
permit the interpretation of nuclear reactions.   The concept of nuclear
temperature is a part of that description.  Statistical methods were
introduced in response to the ``chaotic" neutron and proton spectra revealed
in various reactions. (Of course the term chaos had not been introduced when
these methods were developed.)  Very importantly it was found that energy
averaging over the chaotic spectrum leads to the optical model.  An
illustration of the effect of energy averaging to yield an isobar analog
resonance is shown in Figure~3.  Finally we come to the weak
interactions.  It was in
$\beta$ decay that parity non-conservation was observed and the description of
$\beta$ decay in terms of V-A interaction was determined.  Tests of the
standard model have been made; others are in progress.  Undoubtedly each of
you would have items which you would add to this list.

We emphasize that the above phenomena and concepts apply to a small system
with a maximum number of a few hundred particles.  Do these show up in other
small systems?  The answer is yes, and a most striking example is given
by mesoscopic physics - Consider metallic clusters for example.  These are
stable combinations of metallic atoms (Na, (the most deeply bound systems) K,
Pb) whose number can range from tens to thousands.  It is found that the
number of atoms in the most abundantly produced clusters form a set of magic
numbers, (see Figure~4) which are conveniently understood as the bound
states of electrons in a mean field, called jellium, which can be taken to
have the Woods-Saxon shape.  A deformed potential leads to better agreement
with experiment.(see Figure~5)  Giant resonances are seen. (see
Figure~6)  And a number of phenomena exist which can be readily
understood when references is made to nuclear analogs.  Of course the nature
of the clusters do change when the number of atoms is relatively large.  A
second example is provided by quantum dots and by very small conductors.  The
phenomena in these cases is very similar to that exhibited by ``chaotic"
nuclear spectra (see Figure~7) and have been treated in an identical
fashion making use of the appropriate nuclear theoretical developments.
Further examples can be cited.  The point to be emphasized is that small
systems exhibit common phenomena -- moreover these occur in spite
of the fact that the interparticle forces are very different in character.
Thus the researches in Nuclear Physics leads to results which can be
characterized as universal in that with suitable modifications they apply to
small systems generally.

We turn now to the future.  It should be emphasized that there is continuing
research in all the areas mentioned above as you can see by simply looking at
the conference program.  But what we shall devote most of the talk to is the
very exciting prospect presented by the three new systems now under study by
nuclear physicists.  There is the study of radioactive beam nuclei which are
far from the stable valley.  What are the properties of nuclei which have an
excessive number of neutrons (e.g.
$^{11}$L) or an excessive number of protons?  The next two systems I will
mention involve properties of QCD, (quantum chromodynamics).  The first is the
study of the structure of the hadrons.  The nucleon for example is now seen to
be a many body system; the particles involved are up and down quarks,
sea quarks, that is strange quarks and anti quarks of the up, down and
strange variety and gluons. The interactions are strong
and the kinematics are relativistic.  A matter of great interest is the effect
of the nuclear medium on the internal structure of the nucleon.  It raises the
general question of how to incorporate the complex internal structure of a
particle into a description of a bound system of such particles.  Or if one
considers the nucleus to be a collection of quarks and gluons how do these
condense into the nucleons forming the usual description of the nucleus?
Still a third aspect of this same problem is how to reconcile the boson
exchange models of nuclear forces with the quark-gluon models.  There is the
question of how strange particles behave in nuclei.  What are the
baryon-lyperon interactions?

We
turn next to the quark-gluon plasma.  This is short hand for the state of
nuclear matter under high density and/or high temperature (i.e. highly
excited).  We will be able to trace the behavior of nuclear matter as density
and/or temperature are increased from low to high values.  This will be
accomplished by the study of the collision of heavy ions from low to
ultra-relativistic energy.  The major issues at the ultra relativistic end
are:  Under conditions of high density and temperature will we form a region
in which quarks and gluons are deconfined?  Will chiral symmetry be restored?
These are the three or four
essentially new areas which provide opportunities for productive and
significant research by nuclear physicists.  In addition one should also
consider the more established sub-fields which continue to expose new
phenomena and new puzzles and importantly provide the intellectual support
needed for the exploitation of the newer initiatives.  New concepts, new
structure of ``universal" importance will be developed within these sub-fields.

\section{The Nucleon}
Let us now discuss these ``new" scientific
opportunities.  We begin with the nucleon.  The model of the nucleon as
consisting of three constituent quarks, in spite of its many successes turns
out to be too simplistic.  It was hoped that this model was a mean field model
which involved an average over the quark-quark, quark-gluon and gluon-gluon
interactions.  That hope in some approximation may be correct but is  now
clear that it is not capable of fully describing the hadron.  One must take
into explicit account the degrees of freedom represented by gluons and
quark-antiquarks, the so called sea quarks.

The principal experimental tools involve electron, or more generally
lepton-nucleon scattering, and proton-proton scattering.  Importantly both
projectiles and targets can be polarized.  The results are expressed in
terms of structure functions including spin structure functions, from which a
description of the hadron will eventually emerge.  The simplest of these are
the form factors obtained by the scattering, elastic and inelastic, of
unpolarized electrons (or leptons) by an unpolarized nucleon.  Considerable
progress has been made although further experiments on the determination of
the neutron form factor are desirable for a precision result.  Spin
structure functions (see Figure~8, 9, 10) have
provided the startling result that 30\% of the nucleon spin resides in the
quarks, -10\% in the strange sea quarks while the remainder must reside in the
gluons which must be polarized or in angular momentum contributions which have
not been included in the calculations.  Another experiment which will bear on
this issue are parity violating electron scattering which can yield the strange
quark contributions to the magnetic moment of the proton.  Another
startling result is the indication that the population of anti up sea quarks
differs substantially from that of the anti down sea quark.  Finally one
should mention the evidence that the nucleon is deformed so that the effective
quark-quark interaction must contain a tensor component.

Investigation of the gluon distribution will be one of the goals of the
($\vec{p},\vec{p}$) collisions.  The protons bring in gluons so that the
reactions like $g + \vec{q} \rightarrow q + \gamma$ and $g+g\rightarrow q+q$.
The first of these yields a direct photon production while the second will
manifest itself in two jets. The production of $W^+$ and $W^-$ reflects the
presence of anti down and anti up quarks respectively.

Internal structure generally guarantees the
existence of excited states and vice versa the properties of excited states and
their excitation can inform us with respect to the internal structure.  Thus a
second phase of the study of the nucleon is the study of excited state --
hadron spectroscopy.  QCD leads one to expect the existence of exotic states,
e.g. mesons whose dominant configuration contains only gluons and/or more
than a single $\bar{q}q$ pair.  From $\bar{p}p$ annihilation at rest one
obtains a rich spectrum of scalar states including the $f_0$(1520) meson, a
glueball candidate.  Great progress has been made in identifying numerous
conventional
$\bar{q}q$ systems.  But the search continues for the predicted glueball and
hybrid states.

At low energies, an effective Lagrangian with consequent chiral perturbation
theory ($\chi$PT) has been derived.  Relevant experimental issues include
$\gamma + p \rightarrow \pi^0 + p$ reaction, the electric polarizability of
charged pions, the spectral shape of the
$\eta\rightarrow 3\pi$ decay.

The next level of study is the hadronic interactions.  The main issue is the
connection of the meson exchange description of nucleon forces with the
quark-gluon degree of freedom.  Or how do the quark-gluon degrees of freedom
manifest themselves in  baryon-baryon scattering?  The nucleon-nucleon sector
has been thoroughly studied but we need information on the hyperon-nucleon
interaction and on the hyperon-hyperon interaction.  What for example are the
properties of the dibaryon system?  Is there a 6 quark system like the one
suggested by Jaffe?  The data is very sparse.  As can be seen from
Figure~11, very little is known regarding the hyperon-nucleon
interaction.?  The best information so far is obtained from the properties of
hypernuclei.

Finally we turn to hadrons in nuclear matter.  Experiments involving the EMC
effect continue.  Beyond that I shall mention only (1) the suspected
phenomenon of color transparency in which very small color singlet objects
which interact weakly with nuclear matter are formed in hard collisions and
(2) the approach to chiral symmetry with increasing density.  Of course
the nuclear gluon distribution as well as that of the anti-quark distribution
have not been adequately studied.

\section{Heavy Ions and the Quark-Gluon Plasma}

There are many fundamental issues connected with heavy ion collisions.  I
will pick on one:  namely can concepts which have been developed to
understand the behavior of macroscopic systems be used to understand phenomena
encountered in heavy ion collisions?  For example is the concept of
temperature valid for these relatively small systems?  And if so what
characterizes thermal equilibrium?  Can one expect to see change of
phase change as the experimental parameters are changed?  Is classical
hydrodynamics valid?  Can transport theory be used?  And what is the impact of
collective forms of motion?  Remember we are dealing with relatively few
constituent particles.  A caveat here.  In Au + Au collisions at the AGS 1600
particles were produced.  Are macroscopic concepts applicable?

A great new frontier is now being opened with the use of relativistic heavy
ion reactions which create systems with high energy and/or matter density.
Will the consequences be deconfinement so that a new form of matter is
generated -- the quark-qluon plasma?  What are the appropriate degrees of
freedom?  What will be the collective forms of motion?  And the most important
question, how will the change from hadronic matter to the quark-gluon
plasma matter takes under the regime of high energy and high density?  For
example it is thought that baryon free matter is generated when two heavy
ions  pass through each other as illustrated in Figure~12.  It is of
better to look at the whole picture as shown in Figure~13.  Here we
can see the possible changes which can occur as the density and/or temperature
($\equiv$ energy density) change.

Parenthetically, these phases of nuclear matter play a role in the history of
our universe, another significant contribution of nuclear to astrophysics.

On this diagram the various paths which might be explored by experiments
located at the AGS at Brookhaven, the SPS at CERN and RHIC at Brookhaven are
sketched.  The AGS provides beams with energies of (11.6A) GeV, the SPS heavy
ion energies are (160A)GeV while at RHIC the colliding beams have energies of
(100A x 100A)GeV.  The first two of these involve fixed target energies so
that the CM energies ar relatively low.  RHIC will be ready for experiment by
1999.

{}From experiments now being conducted at the AGS and SPS one learns of the
existence of a regime in which 1/2 of the nucleons are in excited states.
Thus has been called ``Baryon Resonance Matter".  Another and surprising
result is the high stopping power - greater than predicted.  At the AGS the
two colliding nuclei Au + Au stop each other  leading to baryon densities
thought to be as high as 10$\rho_0$ where $\rho_0$ is the normal nucleon
density.  Strangeness production is enhanced, multiply strange baryons as well
as anti-strange baryons are produced.  The suppression of$J/\psi$ and $\psi'$
increases with increasing transverse energy.  An excess in the production of
dilepton pairs of low and intermediate mass is reported for the collision of
protons and light ion collisions at the SPs.  These results are provocative.
Further experiments are needed.  It will be fascinating to follow their
behavior as the center of mass energy increases to the RHIC maximum of 200A
GeV.  Will there be visible indications of the change from nuclear matter to
the quark-gluon plasma (see Figure~14)?.

Relatistic heavy ion physics presents a very exciting prospect with many
opportunities for discoveries of a fundamental-universal importance.

\section{Radioactive Nuclei}

This research area is concerned with nuclei which are neutron (or proton)
rich.  They are radioactive and approach the neutron (proton) dripline.  The
nuclei in the stable valley number a few hundred whereas the radioactive nuclei
number a few thousands.  As can be seen in Figure~14, in the case of
the stable nuclei the spatial distribution of the neutrons and protons are
substantially identical, the neutron and proton radius are equal to within
roughly 0.1 fm.  For the neutron rich nuclei (e.g. $^{11}$Li, $^8$He) outside
of the stable valley this is no longer the case (see Figure~16).  The
surplus neutrons are found just outside core when the two neutron separation
energy is of the few MeV.  One then speaks of a neutron skin.  When that
energy is very small certainly less than 1 MeV one finds a spatial neutron
distribution with a very long tail of low density.  This is confirmed by the
momentum distribution which a corresponding narrow peak.  These nuclei are
said to possess neutron halos (see Figure~17).  All halo nuclei except
$^{11}$B, have two neutrons in the last orbital.   This indicates that the
binding of these nuclei is a consequence of the interaction of these two
neutrons implying a strong two neutron correlation maximizing their attractive
potential energy.

One can ask can the properties of these nuclei be obtained from what we know
about the nucleon-nucleon interaction from the properties of the stable
nuclei?  What collective forms of motion will they exhibit?  It is already
surmised that they are deformed so that there should be the corresponding
rotational levels.  Super and hyper deformation are anticipated (see
Figure~18).  It will also be possible to probe the super-heavy
region.  In the case of the halo nuclei crudely one expects an oscillation of
the two halo neutrons against the core nucleus as illustrated in
Figure~19.  This it is believed has been observed by in photo
disintegration where one finds a ``soft" El mode which can be described as a
resonance.  The radioactive nuclei are polarized permitting the measurement of
their magnetic moments.

Research in sub-field is just beginning.  Only the surface has been
scratched.  What shall we find when we dig deeper?

\section{Weak Interactions}

The study of the weak interactions at the present time has several facets.
One is the study of the properties of the various neutrinos $\nu_e$,
$\nu_{\mu}$, $\nu_t$.  Do they have mass?  Can an oscillation among them be
induced?  Is there a Majorana type neutrino?  Experiments which hear on that
question include double $\beta$ decay, the $\beta$ decay of $^3$H, the solar
neutrino problem, and elastic neutrino scattering.  Long baseline experiments
are planned in which neutrinos are observed after having traveled a large
distance to see whether or not the neutrinos have changed character.

A second front is testing the standard model.  K decay is an important
example.  Another possibility for studying CP and CPT violations lies in the
decay of the $\phi$ meson.  The study of the CKM matrix continues.

In a third front the weak matrix elements in the baryon-baryon sector are
revealed in the decay of the $\Lambda$ in a hypernucleus via $\Lambda
+N\rightarrow N+N$.  Another possible process is $\Lambda + N
\rightarrow N+N+N$.  Recently it was found that parity non conservation could
be detected by the scattering of low energy neutrons by nuclei (see
Figure~21).  The present experiments clearly reveal the existence of
doorway states which were postulated some thirty years ago to explain the
dependence of the neutron strength function on the mass number.  Finally and
most importantly the electron accelerator facilities provide a means for
studying the standard model and in addition provides a probe which through the
parity non conserving interaction can yield information on the weak matrix
elements at the quark level.

\section{Nuclear Theory}

Nuclear theory's function is to act as a guide, as
an interpretator and finally as a synthesizer.  As new areas are investigated,
theory needs to study and suggest the incisive experimental signals which
provide probes of the system under investigation.  On the other hand it must
attempt to characterize the systems by calculations from first principles.
These are of course not independent endeavors.  Of course experiment is a
collaborator and often an instigator.

The computer has become an important  tool for theory.  Using statistical
methods, and such formalisms as the Fadeev theory, the ground state energies
of the light nuclei have been successfully calculated as illustrated by the
table below.  This provides unique information on the nucleon-nucleon
potential.  Only the two body potential is used as the effect of the
many-body nuclear forms can be shown to be small as a consequence of chiral
symmetry.  There is hope that statistical methods can be extended in such a
fashion as to make calculation of the states of heavy nuclei possible.
Clearly as we go more completely into the computer age, highly sophisticated
methods will be developed and problems now regarded as totally intransigent
will be routinely solved.

\vspace*{0.2in}

\begin{tabular}{llllll}
Nucleus ($J^{\pi}$) & $^2H(1^+$) & $^3 H(\frac{1}{2}^+$) & $^4 He(0^+$) & $^5
He(\frac{3}{2}^-$) & $^6 Li(1^+$) \\
Expt (MeV) & -2.22 & -8.48 & -28.3 & -27.2 & -32.0 \\
Theory (MeV) & -2.22 & -8.47(2) & -28.31 & -26.5(2) & -32.4(9)
\end{tabular}

\vspace*{0.2in}

Another example of the use of computers is QCD.  QCD is a non-linear theory
which makes it a fortiori difficult.  Progress has been made on a number of
fronts.  One is lattice gauge theory used to study nucleon and meson
structure.  It from such calculations that we have learned of the importance
of instantons.  A second is the use of sum rules, generally derived from QCD,
which permit the extraction of the contributions of quarks, sea quarks,
strange quarks and gluons from experimental data.  A third approach is know as
chiral perturbation theory.  It is an effective theory consistent with QCD
which is presumed to apply at low energies.  It makes use of the fact that the
Goldstone bosons decouple at zero energy so that an expansion in powers of the
energy.

In the case of heavy ion physics the computer makes it possible to calculate
hadronic and partonic cascades.  The former has been successfully used to
understand the collision of heavy ions at the AGS and the SPS.  At
intermediate energies a one body transport equation (LBUU) is very
successful.  At relativistic energies lattice-gauge calculations have
been used to explore the properties of the quark-gluon plasma.

The quantative consequences of QCD remains unsolved, although the
approaches described above have made some progress toward this goal.  A fully
relativistic theory which can simultaneously describe structure and reactions
is still a considerable distance away.  It is the major problem certainly of
nuclear physics, but beyond that, of theoretical physics generally.  But with
the coming of age of computers, and the construction of high energy
accelerators it may become possible with the combined efforts of
experimentalists and theorists to find a solution.

Symmetry continues to be of great interest.  We have already noted the
effective use of chiral symmetry.  Symmetry is the underlying motif of the
interacting boson model and other algebraic formulisms.  SU(3) symmetry is
used in the derivation of the nucleon sum rules mentioned above.  It is used
for the coupling constants in deSwart's et.al's description of the
baryon-baryon interaction.  A neglected field in my opinion in spite of its
long history, is the study of SU(3) symmetry and its violations.

It is not possible to finish this address without pointing to the
extraordinary increase in technical capabiilty which has occurred in recent
years.  Accelerator physicists have designed and built facilities which
provide nuclear researchers with beams of high quality with a great range in
energy and in particle species -- and importantly these beams can be
polarized.  CW beams make coincidence experiments feasible.  Accompanying
these developments are improvements in detector efficiency and solid angle
coverage permitting a full description of the reaction under study.

\newpage

\begin{center}
{\Large\bf FIGURE CAPTIONS}
\end{center}

Figure~1.~
Contributions of Nuclear Physics.

Figure~2.~
Gamma-ray spectra in $^{152}$Dy obtained by summation of gates set on
most members of the superdeformed band.

Figure~3.~
The top figure is a high resolution cross section for the reaction
$^{92}M_0(p,p)$.  The lower curve is a poor resolution cross
section.

Figure~4.~
Sodium cluster abundance spectrum:  (a) experimental (after
Knight {\it et al.}, 1984); (b) dashed line, using Woods-Saxon potential (after
Knight {\it et al.}, 1984); solid line, using the ellipsoidal shell
(Clemenger-Nilsson) model (after de Heer, Knight, Chou, and Cohen, 1987).

Figure~5.~
Fission dissociation energies, $\Delta_{26,P}$ for the
doubly cationic $K^+_{26}$ cluster as a function of the fission channels $P$.
Solid dots:  Theoretical results derived from the SE-SCM method.  Open
squares:  Experimental measurements.  Top panel:  The spherical model compared
to experimental data.  Middle panel:  The spheroidal model compared to
experimental data.  Lower panel:  The ellipsoidal model compared to
experimental data.

Figure~6.~
Photoelectron spectra of small alkali cluster anions.

Figure~7.~
Comparison of magnetoresistance structure from simulation
and experiment.  The experimental system is a Au-Pd wire of length 7900 $\AA$
and width 500 $\AA$.  The simulated wire is 400 by 40 sites, with W=0.6 and E=
0.2.

Figure~8.~
Structure Functions.

Figure~9.~
The spin structure function $g_1^\beta(x)$ for the proton.

Figure~10.~
The ordinate represents increasing orders of approximation.

Figure~11.~
Total cross-sections for baryon-baryon scattering.

Figure~12.~
The collision of relativistic heavy ions.

Figure~13.~
Phase Diagram.

Figure~14.~
Signatures of Quark-Gluon Plasma.

Figure~15.~
Chart of the Nuclei.

Figure~16.~
HF potential for $^{16}O$ isotopes.

Figure~17.~
The ratio of the proton and the neutron density is almost constant
everywhere in a stable nucleus.  However the neutron skins develops when
more neutrons are added on unstable neutron rich nuclei.  Then neutron
halos are formed in nuclei near and on the dripline.  The proton skin
may also be formed in proton rich unstable nuclei.

Figure~18.~
Various exotic shapes of nuclei.  Exotic orbitals
apears in the region far from the stability line provides many types of
deformations.

Figure~19.~
The soft modes of various multi polarities are expected in nuclei with
neutron skin and/or neutron halo.  The resonances provide us a mean to
study the compressibility or other properties of asymmetric nuclear
matter.

Figure~20.~
Magnetic and Quadrupole moments for radioactive nuclei.

Figure~21.~
The parity nonconserving longitudinal asymmetries $P$
for 23 $p$-wave resoin $^{232}Th$.

\end{document}